\documentclass[journal]{IEEEtran}
\usepackage{amscd, amssymb}
\usepackage[mathscr]{eucal}
\newtheorem{theorem}{\bf Theorem}[section]

\newtheorem{lem}[theorem]{\bf Lemma}

\newtheorem{thm}{\bf Theorem}

\newcommand{\eep}{\hfill $\blacksquare$}

\usepackage{epsfig}
\title{\huge Column Weight Two and Three LDPC Codes with High Rates and Large Girths\vspace{-0.5cm}}
\begin{document}
\maketitle

\begin{abstract}\rm{}
In this paper, the concept of the {\it broken diagonal pair} in the
chess-like square board is used to define some well-structured block
designs whose incidence matrices can be considered as the
parity-check matrices of some high rate cycle codes with girth 12.
Interestingly, the constructed regular cycle codes with row-weights
$t$, $3\leq t \leq 20$, $t\neq 7, 15, 16$, have the best lengths
among the known regular girth-12 cycle codes. In addition, the
proposed cycle codes can be easily extended to some high rate column
weight-3 LDPC codes with girth 6. Simulation results show that the
constructed column weight 3 QC LDPC codes remarkably outperform QC LDPC codes based
on Steiner triple systems and integer lattices.
\vspace{.2cm}
\noindent 
\end{abstract}
\vspace{-0.2cm}
\section{{Introduction}}
Low-density parity-check (LDPC) codes~\cite{gal} are the most promising class of linear codes due
to their ease of implementation and excellent performance over noisy
channels when decoded with message-passing algorithms~\cite{sum}.
Based on methods of construction, LDPC codes can be divided
into two categories: random codes \cite{mac} and structured
codes~\cite{protograph}-\cite{bocharova}. Although randomly
constructed LDPC codes of large length give excellent bit-error rate
(BER) performance~\cite{mac}, the memory required to specify the
nonzero elements of such a random matrix can be a major challenge
for hardware implementation. Structured LDPC codes can lead to much
simpler implementations, particularly for encoding.

To each parity-check matrix $H$ of an LDPC code, the Tanner graph TG$(H)$~\cite{tan1} is assigned and 
the girth of the code, denoted by $g(H)$, is defined as the length of the
shortest cycle in ${\rm TG}(H)$. 
Cycles, especially short cycles, in ${\rm TG}(H)$ degrade the performance of LDPC decoders, because they affect the independence of the extrinsic information exchanged in the iterative decoding \cite{mac}.
Accordingly, the design of LDPC codes with large girth is of great interest. 


Cycle codes are a class of LDPC codes with parity-check matrices having fixed column weight-2,  have
shown potential in some applications such as partial response
channels \cite{song3}. Also, it has been shown \cite{cylinder} that designing cycle codes with large girth~\cite{ghol1},~\cite{malm}, especially
in the non-binary setting \cite{n_b_c_codes}, \cite{ncycle1}, is highly beneficial for the
error-floor performance. 

Constructing cycle codes with large girth has been
investigated by several authors. In \cite{cage}, cage graphs were used to construct cycle codes over a wide range of girths and rates. However,
the problem of constructing cage graphs is very challenging and
there is no deterministic approach to constructing arbitrary cages. In addition, in \cite{golomb} the authors used singer perfect difference sets to constructed some non-binary cycle codes with girth 12 and regularity $t=q+1$, $q$ prime power, which achieved the Gallager bound.  In \cite{cylinder}, the authors constructed a particular class of cycle codes with girth $8e$, $e\ge 2$, and rate $1/e$. Subsequently, in \cite{ghol5}, some girth-8 cycle codes were constructed whose parity-check matrices used as the mother matrices of some quasi cyclic (QC) cycle codes with girth 24.

In this paper, some girth-12 cycle codes are presented such that the constructed regular cycle codes with row-weights $t$, $3\leq t \leq 20$, $t\neq 7, 15, 16$, have the best known lengths among known regular girth-12 cycle codes \cite{cage}. Specially, for $t=q+1$, $q$ prime power, the Gallager bound has been achieved for the minimum lengths of the constructed codes. Our construction cause to obtain the memory efficiency in storing the parity-check
matrices of the constructed codes in the decoder. In addition, the parity-check matrices of the proposed girth-12 cycle codes can be extended to some parity-check matrices of column weight three corresponding to some high rate column weight three LDPC codes with girth 6. Simulation results show that the constructed column-weight three QC LDPC codes remarkably outperforms integer lattice and STS based LDPC codes over the additive white Gaussian noise channel.

\section{{Preliminaries and Constructions}}

Let $V=\{0,1,\ldots,m-1\}$ and ${\cal B}=\{B_1,B_2,\ldots, B_b\}$ be a collection containing subsets $B_i\subseteq V$, $1\leq i\leq b$. The {\it incidence matrix} of ${\cal B}$ is an $m\times b$ binary
matrix ${H}=(h_{ij})_{0\le i<m, 1\le j\le b}$, in which $h_{ij} = 1$
iff $i\in B_j$.

\begin{figure*}
\begin{center}
\centerline{\includegraphics[scale=1]{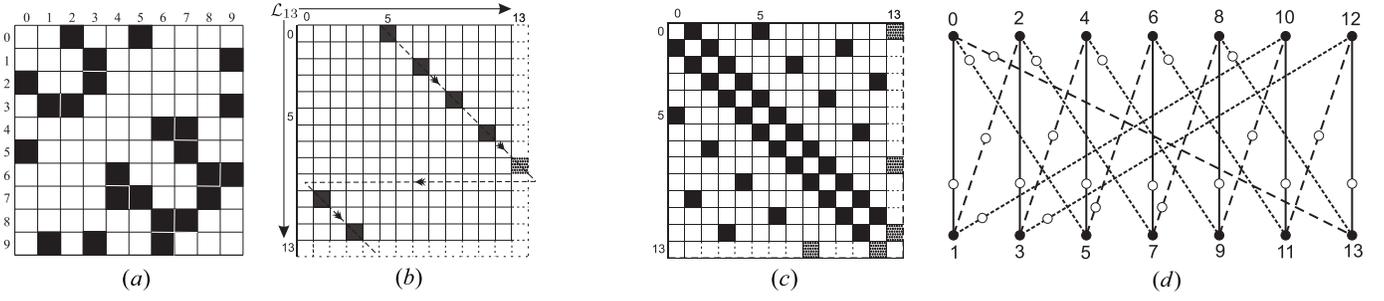}}
\vspace{-.3 cm}
\caption{A random coloring of ${\cal L}_{10}$, lines ${L}_{13}(5)$ and $L_{14}(5)$ in ${\cal L}_{14}$, ${\bf L}_{14}(1, 5, 13)$ and its
Tannar graph, resp. from left to right.}
\label{cage}
\end{center}
\vspace{-1 cm}
\end{figure*}

For a given integer $m$, let ${\cal L}_m$ denote the $m\times m$
square board whose columns (resp. rows) are indexed by
$0,1,\ldots,m-1$ from left to right (resp. up to down) starting from
the most upper-left corner. So, by the square $(i,j)$, $0\le i,j\le
m-1$, we mean the square with row and column indices $i$ and $j$,
respectively. The {\it main diagonal} of ${\cal L}_m$ is defined as
the $\{(i,i), 0\le i\le m-1\}$. By a {\it coloring} of ${\cal L}_m$,
we mean a white-black coloring of the squares so that the main
diagonal squares are white and the black squares are symmetric with
respect to the main diagonal of ${\cal L}_m$, i.e. if the square
$(i,j)$ is black, then the square $(j,i)$ is also black. For
example, a random coloring of ${\cal L}_{10}$ is given in
Figure~\ref{cage}, part (a). Since in an arbitrary coloring of
${\cal L}_m$, black and white squares are symmetric, thus we use
$\{i,j\}$ to denote the squares $(i,j)$ or $(j,i)$. For an arbitrary
coloring of ${\cal L}_m$ with black squares $B_k=\{i_k,j_k\}$, $1\le
k\le b$, let ${\cal B}=\{B_1,B_2,\ldots, B_b\}$ and $H=(h_{p,q})$ be
the $m\times b$ incidence matrix of ${\cal B}$.

Clearly $H$ can be considered as the parity-check matrix of a cycle code with design rate $R=1-m/b$ and block length $b$.  
For example, the incidence matrix $H$ corresponding to the random
coloring of ${\cal L}_{10}$, shown in part (a) of Figure~\ref{cage},
is as follows, which can be considered as the parity-check matrix of
a cycle code with girth 12 and design rate 0.25.
\vspace{-.1cm}
\begin{equation}
H={\tiny \left( \begin {array}{cccccccccccc}
   1&1&0&0&0&0&0&0&0&0&0&0
\\ 0&0&1&1&0&0&0&0&0&0&0&0
\\ 1&0&0&0&1&0&0&0&0&0&0&0
\\ 0&0&1&0&1&1&0&0&0&0&0&0
\\ 0&0&0&0&0&0&1&1&0&0&0&0
\\ 0&1&0&0&0&0&0&0&1&0&0&0
\\ 0&0&0&0&0&0&1&0&0&1&1&0
\\ 0&0&0&0&0&0&0&1&1&0&0&1
\\ 0&0&0&0&0&0&0&0&0&1&0&1
\\ 0&0&0&1&0&1&0&0&0&0&1&0
\end {array} \right)}
\label{eq1}
\end{equation}

$g(H)$ is fully dependant on the coloring of ${\cal L}_m$. Although
increasing the number of black squares in ${\cal L}_m$ increases the
rate, in most cases this increment is accompanied by a reduction of
the girth. Therefore for a fixed $g$, the existence of a high rate
cycle code with girth $g$ will be guaranteed by an appropriate
coloring of ${\cal L}_m$. For example,  an appropriate coloring of
${\cal L}_m$ has been presented in \cite{ghol5} to construct some
high rate girth-8 cycle codes with minimum lengths.
In the sequel, some appropriate colorings of ${\cal L}_m$ are
presented to construct some high rate cycle codes with girth 12.

We begin with some notations and definitions. The {\it pandiagonals}
of ${\cal L}_m$ are those diagonal segments that are parallel to the
main diagonal. Two pandiagonals that together contain $m$ squares
are called a {\it broken diagonal pair}. Now, let $m\geq 14$ and $p$
be an odd positive integer less than $m$. For even $m$, we use
$L_m(p)$ to denote the broken diagonal pair in ${\cal L}_m$
containing alternative white and black squares starting from $(0,p)$
such that the square $(0,p)$ is black. For odd $m$, $L_m(p)$ is
obtained from $L_{m+1}(p)$ by removing the last row and column of
${\cal L}_{m+1}$.  It can be easily seen that, for even $m$,
$L_m(p)$ contains black squares $\{(2j,p+2j~(\bmod m)): 0\leq j\leq
\frac{m}{2}\}$ and for odd $m$,
$L_m(p)=L_{m+1}(p)\setminus\{(m-p,m)\}$. For example,
$L_{14}(5)=\{(0,5),(2,7),(4,9),(6,11),(8,13),(10,2),(12,4)\}$ and
$L_{13}(5)=L_{14}(5)\setminus \{(8,13)\}$, which are shown in Figure
\ref{cage} part (b). Also we denote by ${\bf L}_m({p})$ the set
$\{\{i,j\}: (i,j)\in L_m({p})\}$.

For positive integers $m$ and $t$, $t<m$, we define a {\it
$(m,t)-$vector} $\bf v$, as a length-$t$ vector ${\bf
v}=(v_1,v_2,\ldots,v_t)$ such that $v_1<v_2<\cdots <v_t$ are
positive odd integers less than $m$. For a given $(m,t)-$vector
${\bf v}=(v_1,v_2,\ldots,v_t)$, let ${\bf L}_m({\bf
v})=\bigcup_{i=1}^t{\bf L}_m(v_i)$. It is noticed that the oddness
of $v_i$'s, $1\leq i\leq t$, implies that $|{\bf
L}_m({v_i})|=\lfloor \frac{m}{2}\rfloor$. Also,   ${\bf
L}_m(v_i)\cap {\bf L}_m(v_j)=\emptyset$, $i\ne j$, and so $|{\bf
L}_m({\bf v})|=t\lfloor \frac{m}{2}\rfloor$. Obviously, ${\bf
L}_m({\bf v})$ defines a coloring of ${\cal L}_m$ with $2t\lfloor
\frac{m}{2}\rfloor$ black squares. As an special case, the proposed
coloring in \cite{ghol5} is associated to ${\bf L}_m({\bf v})$,
where ${\bf v}=(1,3,\cdots,2\lfloor\frac{m}{2}\rfloor-1)$.

Now, for a fixed $(m,t)$-vector ${\bf v}=(v_1,\ldots,v_t)$, 
let $H_m({v}_i)$ denote the $m\times \lfloor \frac{m}{2}\rfloor$
incidence matrix of ${\bf L}_m({v}_i)$ and $H_m({\bf
v})=(H_m({v}_1), H_m({v}_2), \ldots , H_m({v}_t))$. It is worth
notice that the parity-check matrix $H_m({\bf v})$ simply determined
by $m$ and $\bf v$ and this significantly reduces the complexity of
the decoder for storing the matrix elements. By the following lemma,
it can be seen that the cycle code with the parity-check matrix
$H_m({\bf v})$ is regular or irregular, depending on the parity of
$m$.

\begin{lem}\label{regularity}
For a  $(m,t)$-vector ${\bf v}=(v_1,\ldots,v_t)$, $H_m({\bf v})$ is
the parity-check matrix of a regular cycle code with regularity $t$,
if $m$ is even and $H_m(v)$ is the parity-check matrix of a
irregular cycle code with $t$ rows of weight $t-1$ and $m-t$ rows of
weight $t$, if $m$ is odd.
\end{lem}
\noindent\textbf{Proof. }By the definition of ${\bf L}_m({v_i})$,
$m$ even, every two distinct elements of ${\bf L}_m({v_i})$ have no
intersection. On the other hand, $|{\bf L}_m({v_i})|=\frac{m}{2}$,
therefore each element of $\{0,1,\ldots, m-1\}$ appears exactly once
in the elements of ${\bf L}_m(v_i)$ and so  ${H}_m(v_i)$, $1\le i\le
t$, has row weight 1, which means that ${H}_m({\bf v})$ has
regularity $t$.

By the previous part, for odd $m$, ${H}_{m+1}({\bf v})$ has
regularity $t$. Since ${\bf L}_m(v_i)={\bf
L}_{m+1}(v_i)\setminus\{\{m-v_i,m\}\}$, thus ${H}_{m}(v_i)$ can be
obtained from ${H}_{m+1}(v_i)$ by removing the last row and the
column corresponding to $\{m-v_i,m\}$. Therefore ${H}_{m}(v_i)$ has
$m-1$ rows of weight $1$ and a row (row indexed by $m-v_i$) of
weight 0. Since for each $i\ne j$, $m-v_i\ne m-v_j$,  $H_m({\bf v})$
contains $t$ rows of weight $t-1$ corresponding to the rows indexed
by $m-v_1$, $m-v_2$, $\ldots$, $m-v_t$, and $m-t$ rows of weight
$t$. \eep

It is noticed that for a given $(m,t)$-vector ${\bf
v}=(v_1,\ldots,v_t)$, we may assume that $v_1=1$, because the girth
is invariant under any permutation on the rows (columns) of
${H}_m({\bf v})$. In addition, the design rate of the cycle code
with the parity-check matrix ${H}_m({\bf v})$ is
$1-{m}/{t\lfloor\frac{m}{2}\rfloor}$ which tends to 1, when $t$
increases. In fact, the actual rate is
$r=1-{(m-1)}/{t\lfloor\frac{m}{2}\rfloor}$, because the rank of
${H}_m({\bf v})$ is $m-1$.

Since for odd $m$, ${H}_m({\bf v})$ is obtained from ${H}_{m+1}({\bf
v})$, thus here in after we just consider the case that $m$ is even.
The Tanner graph TG$(H_m({\bf v}))$ consists check nodes
$\{0,\ldots, m-1\}$ and variable nodes $c_{i,j}=\{2j,v_i+2j\pmod
m\}$, $0\le j< \frac{m}{2}$, $1\le i\le t$, corresponding to the
elements of ${\bf L}_m({\bf v})$, where each $c_{i,j}$ connects even
check node $2j$ to odd check node $v_i+2j\pmod m$. As an example,
TG$(H_{14}(1,5,13))$ is shown in Figure \ref{cage}, part (d), in
which variable nodes and check nodes are denoted by white and black
circles, respectively.

\begin{figure}
\begin{center}
\includegraphics[scale=0.62]{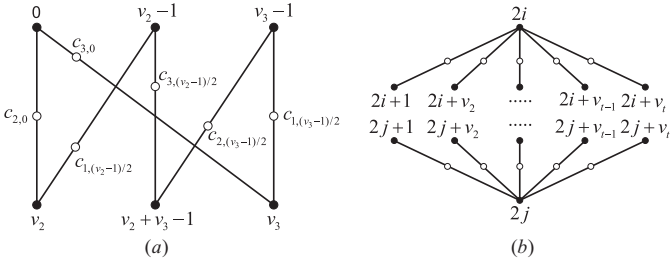}
\end{center}\vspace{-0.5cm}
\caption{\scriptsize (a)~A trivial 12-cycle in TG$(H_m(\bf{v}))$~(b)~Nodes adjacent to $2i$ and $2j$.}\label{fig2}
\vspace{-0.5cm}
\end{figure}

Since the components $v_i$, $1\le i\le t$, of a $(m,t)$-vector ${\bf
v}=(v_1,v_2,\ldots,v_t)$ are distinct, thus TG$(H_m({\bf v}))$ is
free of 4-cycles. On the other hand,  variable nodes in TG$(H_m({\bf
v}))$ connects even check nodes to odd check nodes and so cycles of
lengths 6 and 10 are avoidable in TG$(H_m({\bf v}))$. In addition,
the Tanner graph  TG$(H_m({\bf v}))$ consists the 12-cycle
containing check nodes  $\{0,v_2, v_2-1,v_2+v_3-1,v_3-1,v_3\}$ and
variable nodes
$\{c_{2,0},c_{1,(v_2-1)/2},c_{3,(v_2-1)/2},c_{2,(v_3-1)/2},c_{1,(v_3-1)/2},c_{3,0}\}$,
as shown in Figure \ref{fig2}, part $(a)$. Thus $g(H_m({\bf v}))\le
12$.  Now, in the following theorem, necessary and sufficient
conditions are given which guarantees that $g(H_m({\bf v}))=12$.

\medskip
\begin{thm}\label{main theorem}
Let $m\ge 14$ be even and  $v=(v_1,v_2,\ldots,v_t)$ be a $(m,t)-$vector. Then $g(H_m({\bf v}))=12$ if and
only if for every $k_1,k_2,k_3,k_4$ with $\{k_1,k_2\}\cap\{k_3,k_4\}=\emptyset$, we have
$$(v_{k_1}+v_{k_2})-(v_{k_3}+v_{k_4})\not\in \{0,\pm m\}.$$


%
\end{thm}
\noindent\textbf{Proof. }To show $g(H_m({\bf v}))=12$, it is
sufficient to prove that TG$(H_m({\bf v}))$ is free of 8-cycles. By
the definition, for each $i$, $0\leq i<\frac{m}{2}$, any check node
$z=2i$ in  TG$(H_m({\bf v}))$ is only connected to check nodes
$2i+v_{_{k}}\pmod m$, $1\le k\leq t$, through variable node
$c_{_{k,i}}$, $1\le k\leq t$, respectively. It is easy to see that
any 8-cycle in TG$(H_m({\bf v}))$ must contains exactly two distinct
even check nodes $2i$ and $2j$, for some $i\neq j$, because each
variable node connects an odd check node to an even check node. Now,
set $A_1=\{2i+v_{_{k}}, k=1,\ldots ,t\}$ and $A_2=\{2j+v_{_{k}},
k=1,\ldots ,t\}.$ As shown in Figure \ref{fig2}, part (b), an
8-cycle containing check nodes $2i$ and $2j$ exists if and only if
$|A_1\cap A_2|\geq 2$. Thus, let $|A_1\cap A_2|\geq 2$ and $2i+
v_{_{k_1}}, 2i+ v_{_{k_2}}\in A_1\cap A_2$, $k_1\neq k_2$, which
implies that $2i+ v_{_{k_1}}\equiv 2j+v_{_{k_3}}$ and $2i+
v_{_{k_2}}\equiv 2j+v_{_{k_4}}$, for some $1\le k_3,k_4 \le t$ with
$\{k_1,k_2\}\cap\{k_3,k_4\}=\emptyset$, where congruent relations
are considered in modulo $m$.

This means that $v_{k_1}-v_{k_4}\equiv v_{k_3}-v_{k_4}\pmod m$ or equivalently 
$m|(v_{k_1}+v_{k_2})-(v_{k_3}+v_{k_4})$. 
 But $-(2m-8)\leq (v_{k_1}+v_{k_2})-(v_{k_3}+v_{k_4})\leq 2m-8,$
which implies that $(v_{k_1}+v_{k_2})-(v_{k_3}+v_{k_4})\in \{0,\pm m\}$, a contradiction. This observation, shows that 8-cycles are avoidable in TG$(H_m({\bf v}))$ and this completes the proof. $\hfill \blacksquare$

Now, in the following algorithm we generate $(m,t)$-vectors ${\bf v}=(v_1,v_2,\ldots,v_t)$, such that  $H_m({\bf v})$ has girth 12. In fact, the following algorithm find the smallest $m$ such that a $(m,t)$-vector exists. Using this algorithm, Table I
presents such $(m,t)-$vectors ${\bf v}$ corresponding to some girth-12 cycle codes with row-weight $t$, $3\le t\le 20$
and rate $r$.  
Interestingly, the minimum length $n$, $n=mt/2$, of a regular cycle
code with row weight $t=q+1$, $q$ prime power, and girth $12$
determined by the Gallager bound~\cite{gal} has been achieved by the
codes given in Table I marked by an star. Moreover, for other values
of $t$, $t\neq 7, 15, 16$ the proposed cycle codes have minimum
lengths among the known cycle codes \cite{cage}.

\medskip
{
\noindent $\begin{array}{c} \hline \hbox{{\bf Algorithm.} {\small Generating
$(m,t)$-vectors $v$ with $g(H_m(v))=12$.}}
\\
\hline
\end{array}$
\begin{enumerate}
\item Let $m\ge 14$ be even and $t\geq 3$.
\item Let $k=2$, $A_1=\{1\}$, $A_2=\{3,5,\cdots,m-1\}$ and $(v_1, v_2)\in A_1\times  A_2$ are chosen arbitrary.
\item If $k=1$ then $m\to m+2$ and go to step 2.
\item  Choose $v_k\in A_k$. Define $A_{k+1}$ as the set of all elements $v\in\{v_{k}+2,v_{k}+4,\ldots,m-1\}$ such that for all ${i_1},{i_2},{i_3}\in\{1,2,\ldots,k\}$ we have $v\not\equiv v_{i_1}+v_{i_2}-v_{i_3}\bmod m$ and  $2v\not\equiv v_{i_1}+v_{i_2}\bmod
m.$
\item  If $A_k=\emptyset$, then  set $k\to k-1$, $A_{k} \to A_{k}-\{v_{k}\}$, and go to step 3.
\item  If $k=t$, then go to step 8.
\item  $k\to k+1$ and go to step 4.
\item  Print $v=(v_1,v_2,\ldots, v_t)$ as a solution.
\end{enumerate}
}
\section{Column weight 3 LDPC codes}
One important invariant affecting the performance of an LDPC code is the
column weight of the parity check matrix. 
For maximum-likelihood decoding, LDPC codes with larger
column weight will give better decoding performances.
Therefore, the smallest number of rows that can be added to the parity-check matrix of a cycle code with girth at least 6 to construct a girth-6 column-weight three LDPC code is an interesting problem. In the sequel, we give an approach to construct some girth-6 column-weight three LDPC codes from the proposed cycle codes such that number of added rows is small as possible. We begin with the following simple, but floristic lemma.

\begin{lem}\rm
Let $m\geq 14$ and ${\bf v}=(v_1,v_2,\ldots, v_t)$ be an arbitrary
$(m,t)$-vector. The minimum number of rows that must be added to
$H_m({\bf v})$ to construct girth-6 column-weight three LDPC code is
at least $t$.

\noindent\textbf{Proof. }
Consider an arbitrary row of $H_m({\bf v})=(h_{ij})$ with regularity
$t$ and non zero elements
$h_{r,c_{1}}=h_{r,c_{2}}=\ldots=h_{r,c_{t}}=1$. To avoid 4-cycles in
any extension of $H_m({\bf v})$ to a column-weigh three parity-check
matrix, we need at least $t$ new rows correspond to the column
indexed by $c_1,c_2,\ldots,c_t$.\eep
\end{lem}

Now, we go through the details of the construction. Let $m\geq 14$
and ${\bf v}=(v_1,v_2,\ldots, v_t)$ be an arbitrary $(m,t)$-vector.
For each $i$, $1\leq i\leq t$, set $B_i=\{\{e,o,m+i-1\}: \{e,o\}\in
{\bf L}_m(v_i)\}$ and ${\cal B}_m({\bf v})=\bigcup_{i} {B}_i$. For
example
\noindent ${\cal
B}_{14}(1$,$5$,$13)=\{\{0$,$1$,$14\}$,$\{2$,$3$,$14\}$,$\{4$,$5$,$14\}$,$\{6$,$7$,$14\}$,$\{8$,$9$,$14\}$,$\{10$,$11$,$14\}$,$\{12$, $13$,$14\}$,$\{0$,$5$,$15\}$,$\{2$,$7$,$15\}$,$\{4$,$9$,$15\}$,$\{6$,$11$,$15\}$,$\{8$,$13$,$15\}$,$\{1$,$10$,
$15\}$,$\{3$,$12$,$15\}$,$\{0$,$13$,$16\}$,$\{2$,$1$,$16\}$,$\{4$,$3$,$16\}$,$\{6$,$5$,$16\}$,$\{8$,$7$,$16\}$, $\{10$,$9$,$16\}$,$\{12$,$11$,$16\}\}$.

\begin{table*}
\vspace{-.5cm}
\begin{center}
${\scriptsize\begin{array}{|c|c|c|c|c|c|c|} \hline
m&t&r&r'&n&\hbox{Gallager bound}&v=(v_1,\cdots,v_t)\\
\hline\hline 14&3&0.38&0.24&21^*&21&(1,5,13)\\\hline
26&4&0.52&0.44&52^*&52&(1,5,17,25)\\\hline
42&5&0.61&0.56&105^*&105&(1,11,15,35,41)\\\hline
62&6&0.67&0.64&186^*&186&(1,15,21,25,33,61)\\\hline
96&7&0.71&0.7&336&301&(1,29,51,71,85,89,95)\\\hline
114&8&0.75&0.73&456^*&456&(1,25,29,41,47,61,105,113)\\\hline
146&9&0.78&0.77&657^*&657&(1,13,21,69,95,101,105,129,145)\\\hline
182&10&0.8&0.79&910^*&910&(1,3,13,21,47,53,69,83,107,111)\\\hline
240&11&0.82&0.81&1320&1221&(1,93,105,125,155,159,181,195,223,233,239)\\\hline
266&12&0.83&0.82&1596^*&1596&(1,5,13,49,59,81,87,111,137,151,153,171)\\\hline
336&13&0.85&0.84&2379&2041&(1,39,61,69,75,93,127,171,175,191,217,325,335)\\\hline
366&14&0.86&0.85&2562^*&2562&(1,31,99,103,109,143,157,169,185,193,231,249,345,365)\\\hline
510&15&0.87&0.86&3825&3165&(1,23,27,71,79,109,167,183,233,243,297,391,491,497,509)\\\hline
510&16&0.875&0.87&4080&3856&
(1,21,23,63,67,117,141,147,155,173,245,255,303,315,331,367)\\\hline
546&17&0.88&0.879&4641^*&4641&(1,11,31,69,71,85,147,151,173,179,197,269,303,311,355,367,403)\\\hline
614&18&0.889&0.886&5526^*&5526&(1,5,21,45,107,113,165,167,179,197,261,297,307,335,377,385,411,433)\\\hline
720&19&0.895&0.892&6840&6517&
(1,7,63,65,83,135,173,189,221,233,257,267,369,397,411,419,485,511,515)\\\hline
762&20&0.9&0.898&7620^*&7620&(1,49,61,87,111,143,151,179,209,251,255,325,335,379,413,431,545,551,565,567)\\\hline
\end{array}}$\\\vspace{0.1cm}
{\small {\bf Table I.}
Some $(m,t)-$vectors ${\bf v}=(v_1,\cdots,v_t)$ associated
to $H_m({\bf v})$ with girth 12.\vspace{-0.6cm}}
\end{center}
\label{tab1}
\end{table*}
Let $ {\cal M}_m({\bf v})$ denote the incidence matrix of ${\cal
B}_m({\bf v})$. It is easy to see that $ {\cal M}_m({\bf v})$ can be
considered as the parity-check matrix of a column weight 3 LDPC code
with girth 6, because for each $i\neq j,$ ${\bf L}_m(v_i)\cap {\bf
L}_m(v_j)=\emptyset$ and the new points added to ${\bf L}_m(v_i)$
and ${\bf L}_m(v_j)$ are distinct. Clearly the first $m$ rows of
${\cal M}_m({\bf v})$ have the same regularity as ${\bf H}_m({\bf
v})$ and the new $t$  added rows have regularity $\lfloor\frac{m}{2}
\rfloor$. Therefore ${\cal M}_m({\bf v})$ is always irregular unless
$m$ is even, $t=\frac{m}{2}$ and ${\bf v}=(1,3, \ldots , m-1)$. The
later case, was discussed in \cite{ghol4}, which can be easily
derived from our construction. Moreover, the rate of the constructed
column weight 3 LDPC codes is $r'=1-\frac{t+m-1}{t\lfloor{m}/{2}
\rfloor}$, which tends to one when $t,m$ increases. As shown in
Table I, the rate of the constructed column weight three LDPC codes
derived from the proposed cycle codes, denoted by  $r'$, is close to
the rate of the proposed cycle codes.

Kim et al. \cite{kim} have shown that if the base matrix $H$ has
girth $2g$, then the maximum achievable girth of quasi cyclic LDPC
codes having base matrix $H$ is at least $6g$. Using this fact, the
constructed column weights 2 and 3 LDPC codes can be considered as
the base matrices of some quasi cyclic LDPC codes with girth at
least 36 and 18, respectively. In fact, using a similar approach
posed in \cite{ghol4}, the maximum achievable girth of the
constructed quasi cyclic LDPC codes with column weight 3 is 20.

%
\vspace{-.3cm}
\section{Simulation Results}
In this section, we examine a performance comparison between the constructed column
weight three LDPC codes with large girth, employing the proposed
algorithm in \cite{ghol4}, on one hand, and LDPC
codes with different girths constructed in \cite{ghol5} based on Steiner triple system $STS(9)$ and the 15-points $3\times5$ integer lattice $L(3\times5)$, on the other hand.

\begin{figure}
\begin{center}
\includegraphics[scale=0.47]{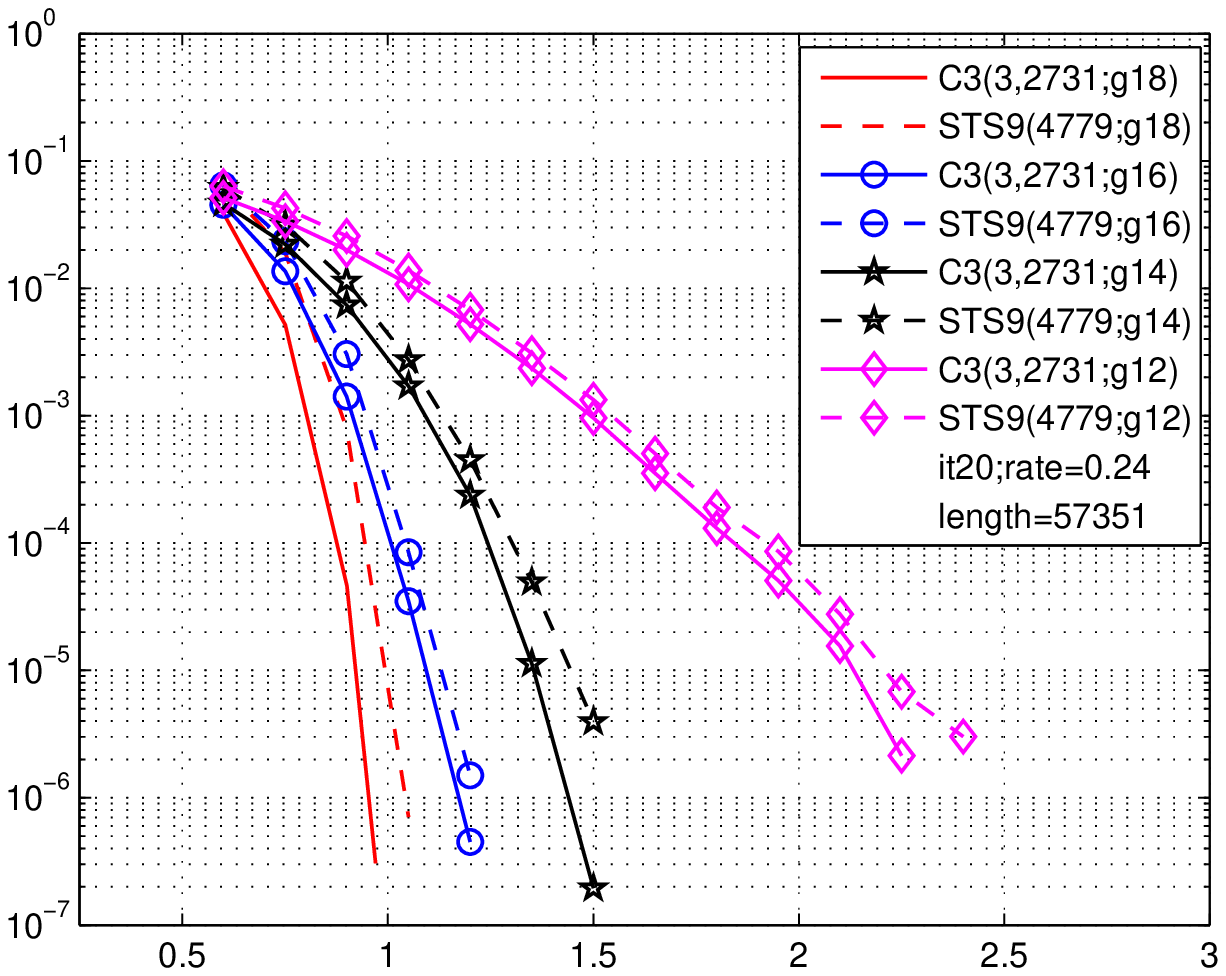}\vspace{-0.25cm}
\includegraphics[scale=0.47]{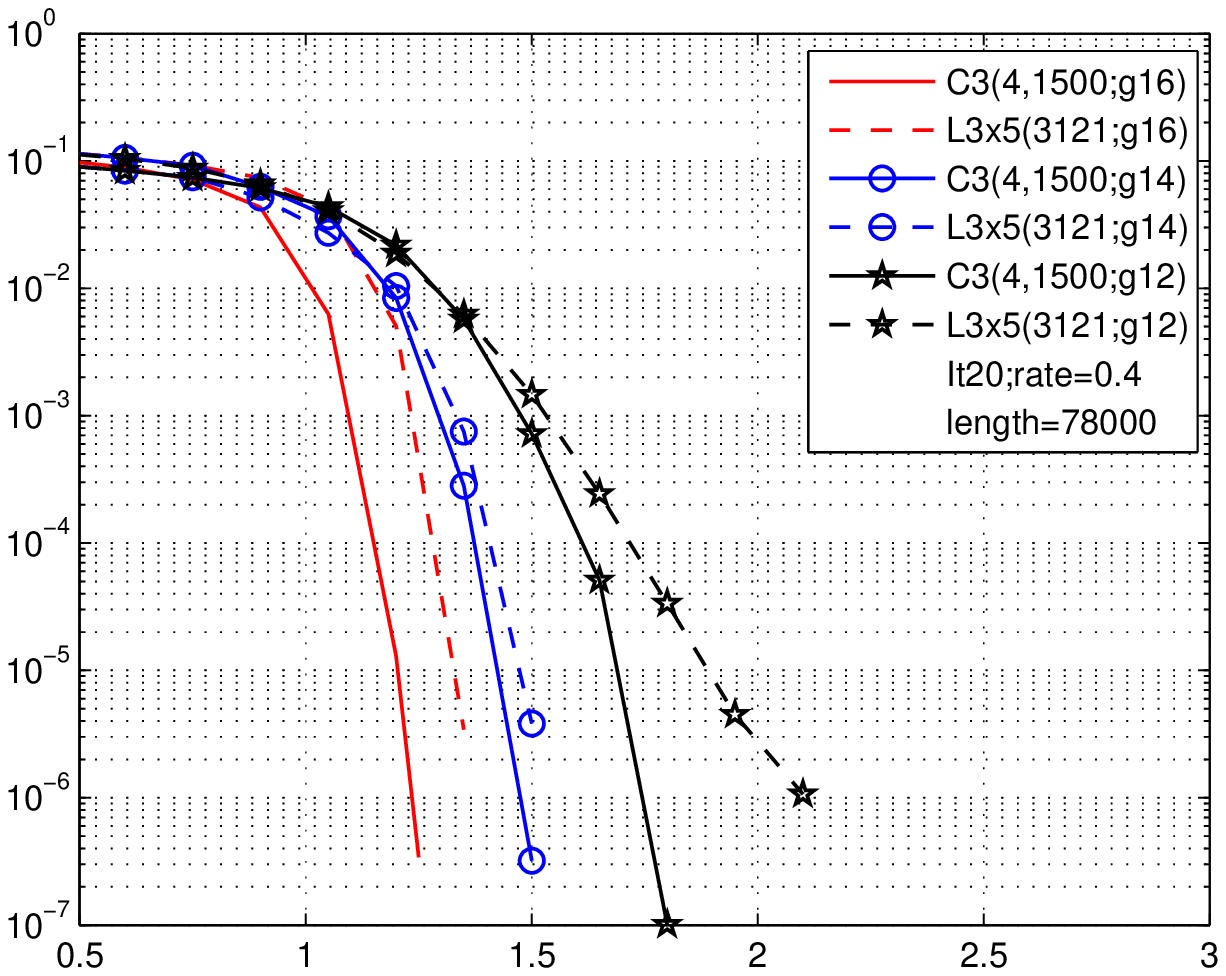}
\end{center}\vspace{-0.5cm}
\caption{Comparisons between the proposed column-weight 3 LDPC codes
with different girths and some well-known codes}\label{sim2}\vspace{-0.5cm}
\end{figure}

In Figure \ref{sim2}, $STS(9)(N;gb)$ and $L3\times5(N;gb)$ are used to denote $STS(9)$ and $L(3\times5)$-based LDPC codes with block-size $N$ and girth $b$, respectively. Moreover, $C3(m,N;gb)$ is used to denote the column-weight three QC LDPC code with girth $b$, block size $N$ which is lifted from the base matrix ${\cal M}_m({\bf v})$. As shown Figure \ref{sim2}, the constructed column-weight three codes with different girths significantly outperform the codes based on $STS(9)$ and $L(3\times 5)$ with the same girth.




\end{document}